\documentclass[10pt,twocolumn,letterpaper]{article}

\usepackage{iccv}
\usepackage{times}
\usepackage{epsfig}
\usepackage{graphicx}
\usepackage{amsmath}
\usepackage{amssymb}

\usepackage{multirow}
\usepackage{subcaption}
\newcommand{\tabincell}[2]{\begin{tabular}{@{}#1@{}}#2\end{tabular}}

\usepackage[pagebackref=true,breaklinks=true,letterpaper=true,colorlinks,bookmarks=false]{hyperref}

\iccvfinalcopy 

\def\iccvPaperID{1671} 

\ificcvfinal\fi
\begin{document}

\title{An Acceleration Framework for High Resolution Image Synthesis}

\author{Jinlin Liu,\ Yuan Yao,\ Jianqiang Ren}

\maketitle

\begin{abstract}
 Synthesis of high resolution images using Generative Adversarial Networks (GANs) is challenging, which usually requires numbers of high-end graphic cards with large memory and long time of training. In this paper, we propose a two-stage framework to accelerate the training process of synthesizing high resolution images. High resolution images are first transformed to small codes via the trained encoder and decoder networks. The code in latent space is times smaller than the original high resolution images. Then, we train a code generation network to learn the distribution of the latent codes. In this way, the generator only learns to generate small latent codes instead of large images. Finally, we decode the generated latent codes to image space via the decoder networks so as to output the synthesized high resolution images. Experimental results show that the proposed method accelerates the training process significantly and increases the quality of the generated samples. The proposed acceleration framework makes it possible to generate high resolution images using less training time with limited hardware resource. After using the proposed acceleration method, it takes only 3 days to train a $1024\times 1024$ image generator on Celeba-HQ dataset using just one NVIDIA P100 graphic card.
\end{abstract}

\begin{figure}[ht]
\begin{center}
  \includegraphics[width=225pt]{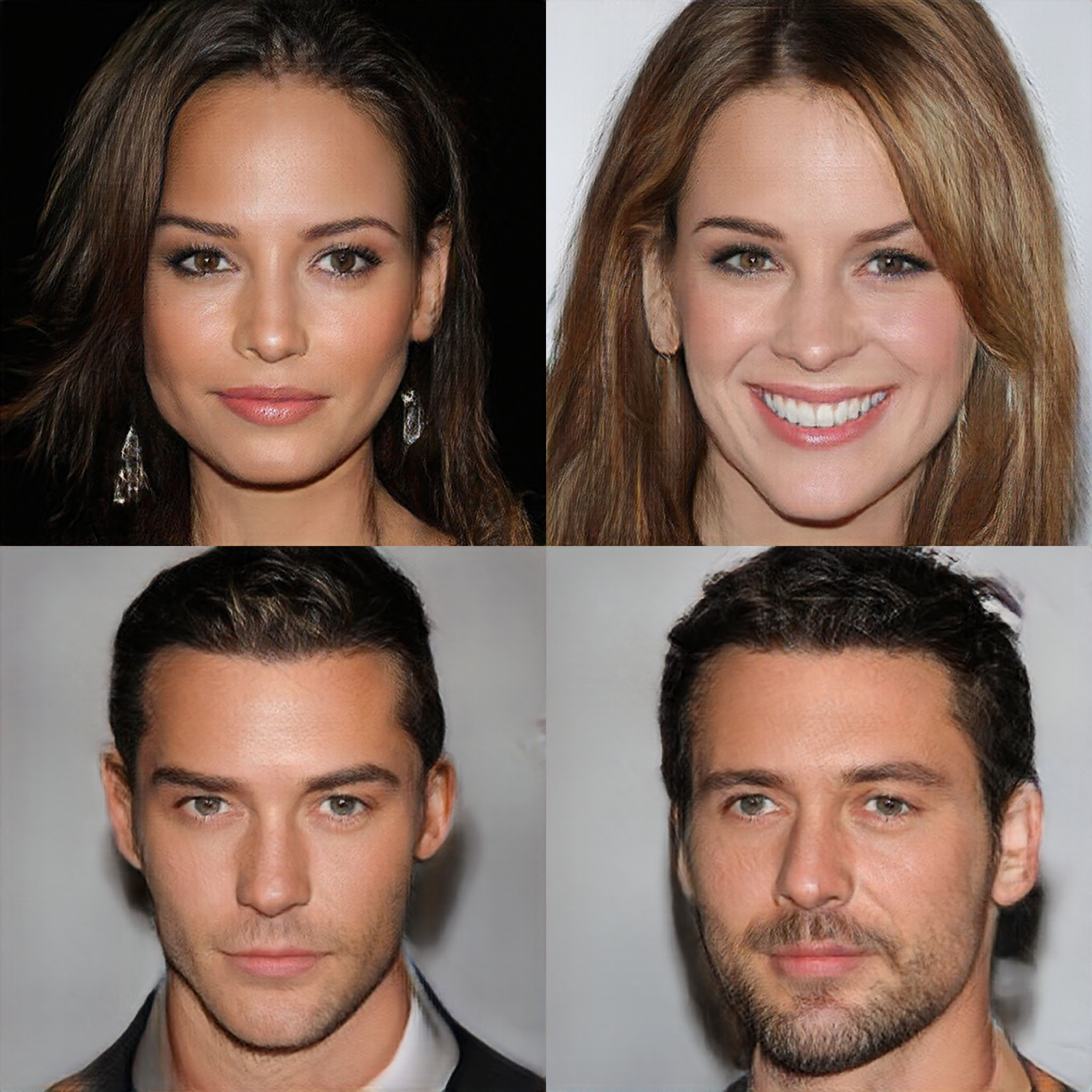}
\end{center}
\caption{Generated $1024\times 1024$ images after training 3 days using only one P100 graphic card by the proposed method.}
\label{fig:tease}
\end{figure}

\section{Introduction}

Generative Adversarial Networks (GANs) are developed to learn the distributions of input data and then generate new samples from the learned distribution \cite{goodfellow2014generative}. Recently, GANs have been applied to lots of tasks and achieve impressive results, such as image enhancement (Demir and Unal 2018; Yu et al. 2018; Nazeri et al. 2019; Ledig et al. 2017; Zhang et al. 2019; Kupyn et al. 2018), high resolution image generation \cite{brock2018large,karras2017progressive,karras2018style,wang2018high}, 3D generation and reconstruction \cite{wu2016learning,smith2017improved,yang20173d}. Training GANs is unstable and sensitive to hyper-parameters \cite{zhang2018self}. Different loss functions and network structures have been developed to train powerful GANs for better quality and stability \cite{arjovsky2017wasserstein,brock2018large,gulrajani2017improved,mescheder2018training,karras2017progressive,salimans2016improved,zhang2017stackgan++,zhang2018self}.

Generation of high resolution images may be difficult for early works, but recent advances in the community have made it possible to generate high quality images even at $512\times 512$ and $1024\times 1024$ resolutions \cite{brock2018large,karras2017progressive,karras2018style}. Even though, we find that training GANs to generate images at high resolutions is both time consuming and computational intensive. High-end graphic cards and long training time are required. Karras et al. \cite{karras2017progressive} reports that it takes NVIDIA DGX-1 with 8 Tesla V100 GPUs and 4 days to train generative networks at $1024\times 1024$ resolution. If only one graphic card is available, it would be 14 days for \cite{karras2017progressive} and 41 days for \cite{karras2018style}. BigGAN \cite{brock2018large} use poweful 128 to 512 cores of a Google TPU V3 Pod so as to scale up GANs.

\begin{figure}[t]
\begin{center}
  \includegraphics[width=225pt]{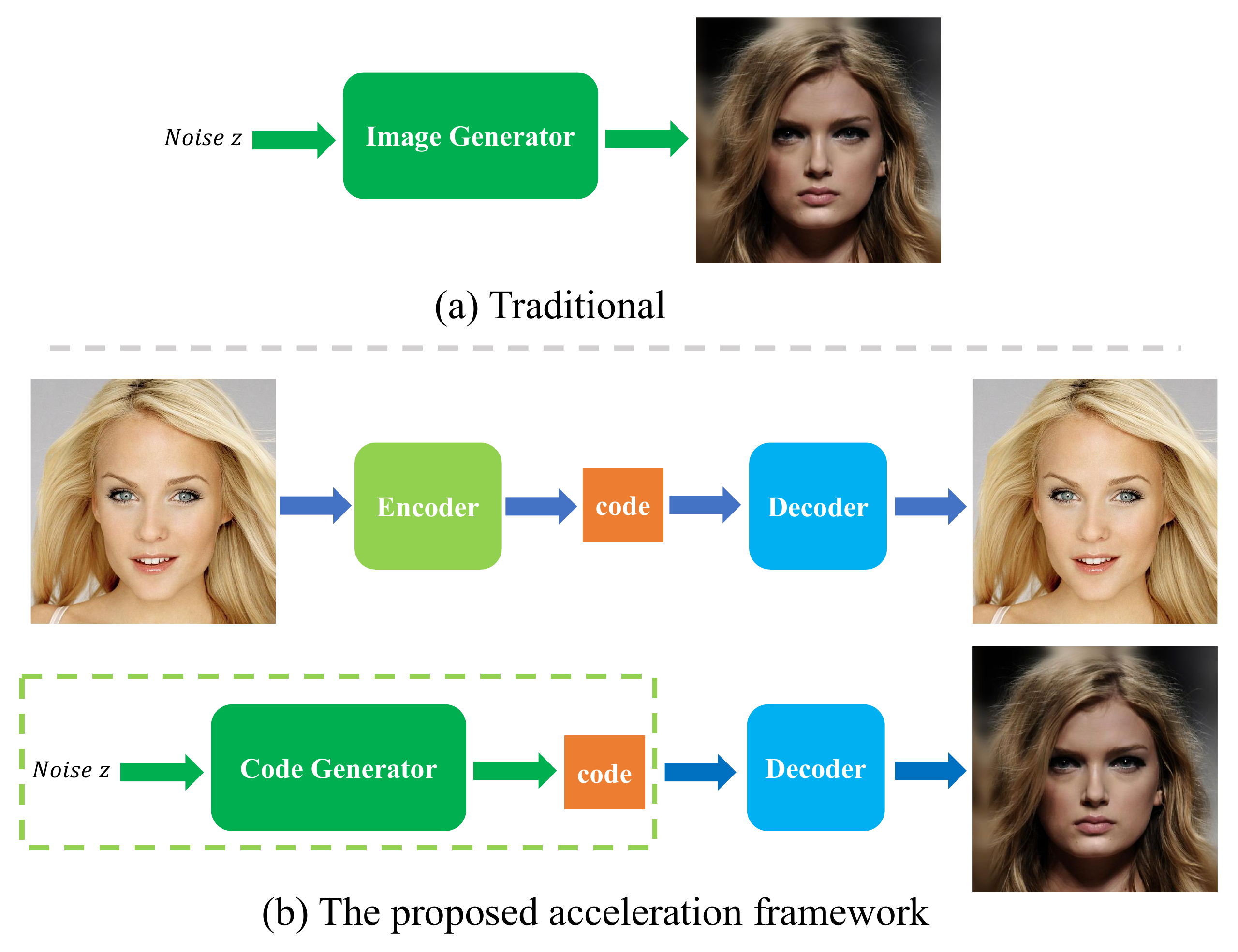}
\end{center}
\caption{The architecture of (a) a traditional image generation structure and (b) the proposed acceleration framework. A traditional structure generates high resolution images from noise directly. The proposed acceleration framework first trains encoder and decoder networks in a supervised way. High resolution images are transformed to small latent codes. Then a code generator is trained to learn the distribution of codes in latent space. Finally, new generated code samples are transformed to high resolution images by the trained decoder network. The proposed framework manages to generate small codes rather then large images, which makes the generation of high resolution images using GANs easier and faster.}
\label{fig:flowchart}
\end{figure}

We propose an acceleration framework in this paper to increase the efficiency of training GANs at high resolution. We manage to generate small codes in latent space instead of large images, as demonstrated in Figure~\ref{fig:flowchart}. Traditional structure generates images from input noise directly, which is challenging to train when image resolution increases. In our framework, we propose a different two-stage way. Encoder and Decoder networks are first trained to transform large images to small latent codes. The generative networks only learns to generate small codes from noise in the latent space. New generated code samples can be easily transformed to high resolution images via the trained decoder network.

The proposed acceleration framework increases the training efficiency for two reasons. First, the latent codes are times smaller than the original large images, which means that less layers and smaller feature maps are in both generator and discriminator networks. Second, though additional encoder and decoder networks have to be trained in the first step, it is easy to train them fast using low resolution images, as both networks are fully convolutional networks. For different resolutions, the encoder and decoder networks only need to be trained once. As a result, in the proposed acceleration framework, we manage to train the networks without feeding in high resolutions images in every training step.

Finally, we build an traditional image generative network and accelerate it using the proposed framework. Experimental results are promising. The training speed is times fast for different resolutions. In addition, the quality of generated samples after acceleration are well improved, benefiting from the better stability of learning the distribution of small codes in latent space than large images.

\section{Related Work}
Training high resolution GANs is challenging and vulnerable to suffer from gradient problems, as the discriminator is easy to distinguish fake images from real images \cite{odena2017conditional}. To train high resolution GANs stably, various techniques have been proposed. Gulrajani et al. \cite{gulrajani2017improved} used gradient penalty to avoid gradient problems and improved the stability in training. Spectral normalization techniques were used by \cite{miyato2018spectral} to stabilize the training of discriminator networks. Zhang et al. \cite{zhang2018self} introduced self-attention module to their network, which enables the network to model long range, multi-level dependencies across image regions. Self-attention mechanism helps improve the details and quality. Kerras et al. \cite{karras2017progressive} and \cite{karras2018style} trained the network progressively, starting from low resolution and growing to large resolutions. Both the generator and discriminator networks grow during the training process which increases training stability significantly. They manage to generate images at $1024\times 1024$ resolution with high fidelity. \cite{brock2018large} and \cite{brock2018large} used powerful computational resource and scaled up GANs using very large batch size and parameter numbers.

The proposed method accelerates the training of GANs without modifying either the network structures or the loss function. We train an additional encoder and decoder network to convert the image generation to the small latent code generation. Variational Autoencoder (VAE) \cite{kingma2013auto} also trained encoder and decoder networks to generate images. During training the encoder and decoder networks, VAEs try to force the code in latent space to obey some kind of distribution (i.e. standard normal distribution). Full resolution images are used in every training step. In the proposed framework, the encoder and decoder networks are trained in a independent way using low resolution images. The parameters of the encoder and decoder networks are fixed after training. Training images are then transformed to latent codes. A generative network is trained to learn the distribution of the latent codes. The proposed method does not use full resolution images in every training step. Both of the target and the training manner of the proposed method are different from VAE methods.

\begin{figure*}[h!]
\begin{center}
  \includegraphics[width=450pt]{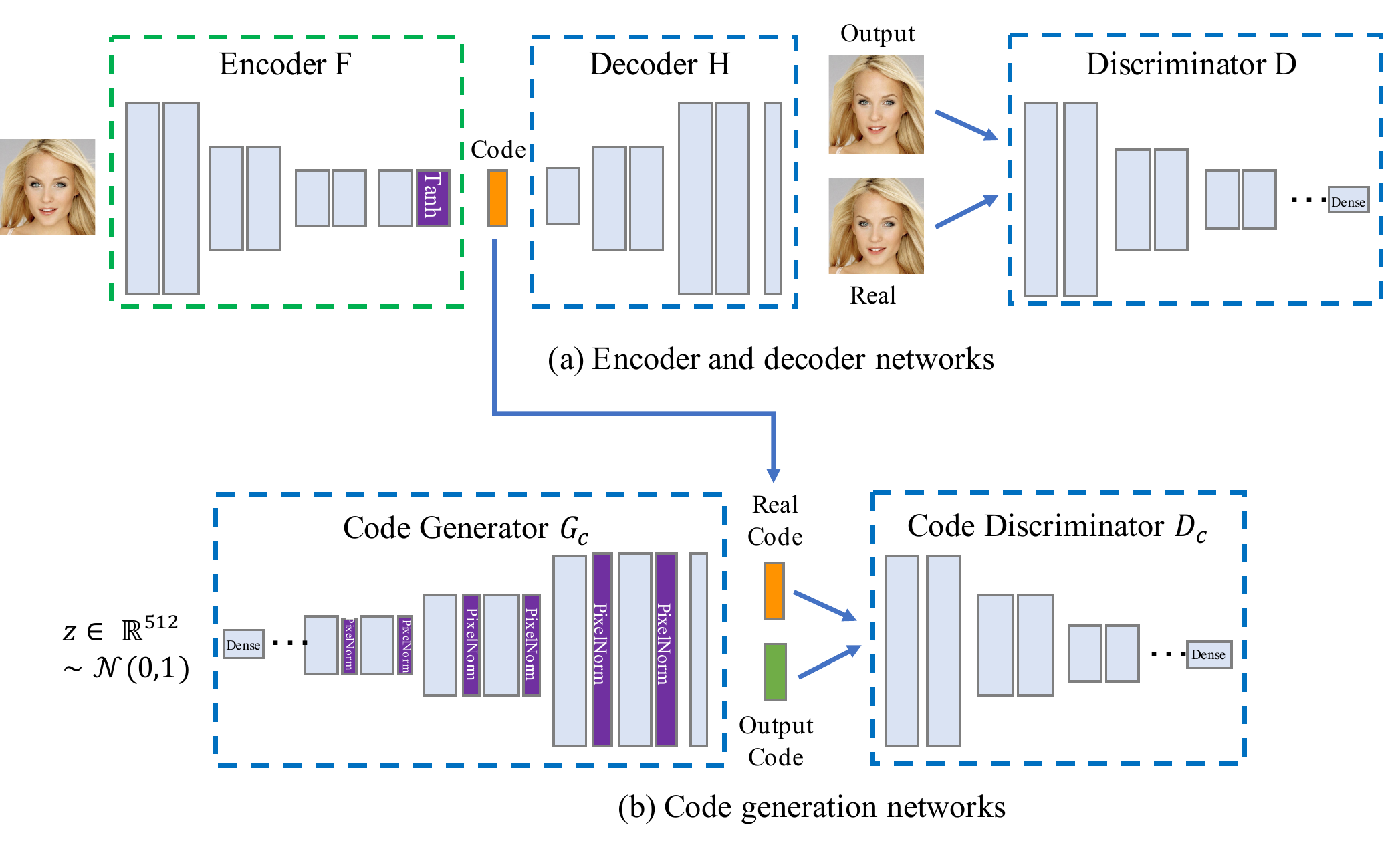}
\end{center}
\caption{The architecture of the encoder and decoder networks and the code generation networks.}
\label{fig:network}
\end{figure*}

\section{The Acceleration Framework}
The key of the proposed accelerating framework for high resolution generation lies in generating small codes in latent space instead of large images in image space. We first train an encoder and decoder network. The encoder network transforms images to latent codes, and the decoder network transforms codes back to images. Then GANs are trained to generate codes in latent space, which can be transformed to images easily by the trained decoder network. The proposed acceleration framework is demonstrated in Figure~\ref{fig:flowchart}.

\subsection{Encoder and Decoder Networks}
The framework of the proposed encoder and decoder networks is demonstrated in Figure~\ref{fig:network}(a). The whole framework is composed of three parts: an encoder network $F$, a decoder network $H$ and an image discriminator network $D$. There are several downsampling operation in the encoder network to transform input images $x$ to latent codes  $c_x$. The decoder network contains upsampling operations to convert latent codes back to images. We add tanh operation at the end of the encoder network, to force the value of the latent codes in [-1,1].

The objective is to minimize the reconstruction error. We adopt $L_1$ loss function here.
\begin{equation}
\begin{aligned}
\label{eqn:01}
\mathcal{L}_{r}=|H(F(x))-x|_1,
\end{aligned}
\end{equation}
where $x$ is the input image. To improve the reconstruction quality, we add an adversarial loss by introducing an image discriminator network. The image discriminator $D$ predicts whether images are real or not. The adversarial loss for the encoder and decoder networks is defined as,
\begin{equation}
\begin{aligned}
\label{eqn:02}
\mathcal{L}_{adv}=\mathbb{E} [-\log (D(H(F(x))))] ,
\end{aligned}
\end{equation}

The total loss for training the encoder and decoder networks is,
\begin{equation}
\begin{aligned}
\label{eqn:03}
\mathcal{L}=\mathcal{L}_{r} + \alpha \mathcal{L}_{adv},
\end{aligned}
\end{equation}

To train the image discriminator, we use the non-saturating loss.
\begin{equation}
\begin{aligned}
\label{eqn:04}
\mathcal{L}_{D}=\mathbb{E} [-\log (D(x) - \log (1-D(H(F(x))))].
\end{aligned}
\end{equation}

\subsection{Code Generation Networks}
The encoder and decoder networks can transform input large images to smaller latent codes and inverse latent codes back to images. Thus, to generate new images from random input noise, we only need to generate small size of codes in latent space. High resolution images $x$ in the dataset are all transformed to latent codes $c_x$ by the encoder network and only those codes are used in the code generation process. In this process, we aim at learning the distribution of the latent codes corresponding to high resolution images.

Though the size of latent codes is different from normal RGB images, the structure of either the generator or the discriminator does not require special customization. The codes and images are both three dimensional matrix and the value range of the codes is $[-1,1]$ from the encoder network, which are exactly the commonly taken in range of generator and discriminator networks. Therefore, the structure of our code generation networks is nearly the same with image generation networks, except that the dimension of the output is the size of latent codes. The architecture of the proposed code generation networks is displayed in Figure~\ref{fig:network}(b), the structure of the code generation networks is identical with a simple image generator and discriminator.


To train the proposed code generation network, we use the WGAN-GP loss function \cite{gulrajani2017improved}, which has been proved to improve stability for image generations. The loss function contains the Wasserstein GAN loss and the gradient penalty loss. The loss function used to train the generator $G_c$ is,
\begin{equation}
\begin{aligned}
\label{eqn:05}
\mathcal{L}_{G_c}=\mathbb{E}_{z\sim p_z} [-D_c(G_c(z))].
\end{aligned}
\end{equation}

The loss function for the code discriminator $D_c$ contains two parts. The first part is Wasserstein loss,
\begin{equation}
\begin{aligned}
\label{eqn:06}
\mathcal{L}_{WGAN}= \mathbb{E}_{z\sim p_z, x\sim p_{data}} [-D_c(c_x) + D_c(G_c(z))],
\end{aligned}
\end{equation}
where $c_x$ is the latent code corresponding to image $x$, i.e. $F(x)$. The second part is the gradient penalt for random smaple $y$,
\begin{equation}
\begin{aligned}
\label{eqn:07}
\mathcal{L}_{gp}=\lambda \mathbb{E}_{c_x\sim p_{c_x}} [(\left\|\nabla D(y)\right\|_2-1)^2].
\end{aligned}
\end{equation}
In experiments, we set $\lambda=10$. The total loss function for our code discriminator is,
\begin{equation}
\begin{aligned}
\label{eqn:08}
\mathcal{L}_{D_c}=\mathcal{L}_{WGAN} + \mathcal{L}_{gp}.
\end{aligned}
\end{equation}

\subsection{Implementation Details}
\label{section:detail}
\begin{figure}
\begin{center}
  \includegraphics[width=225pt]{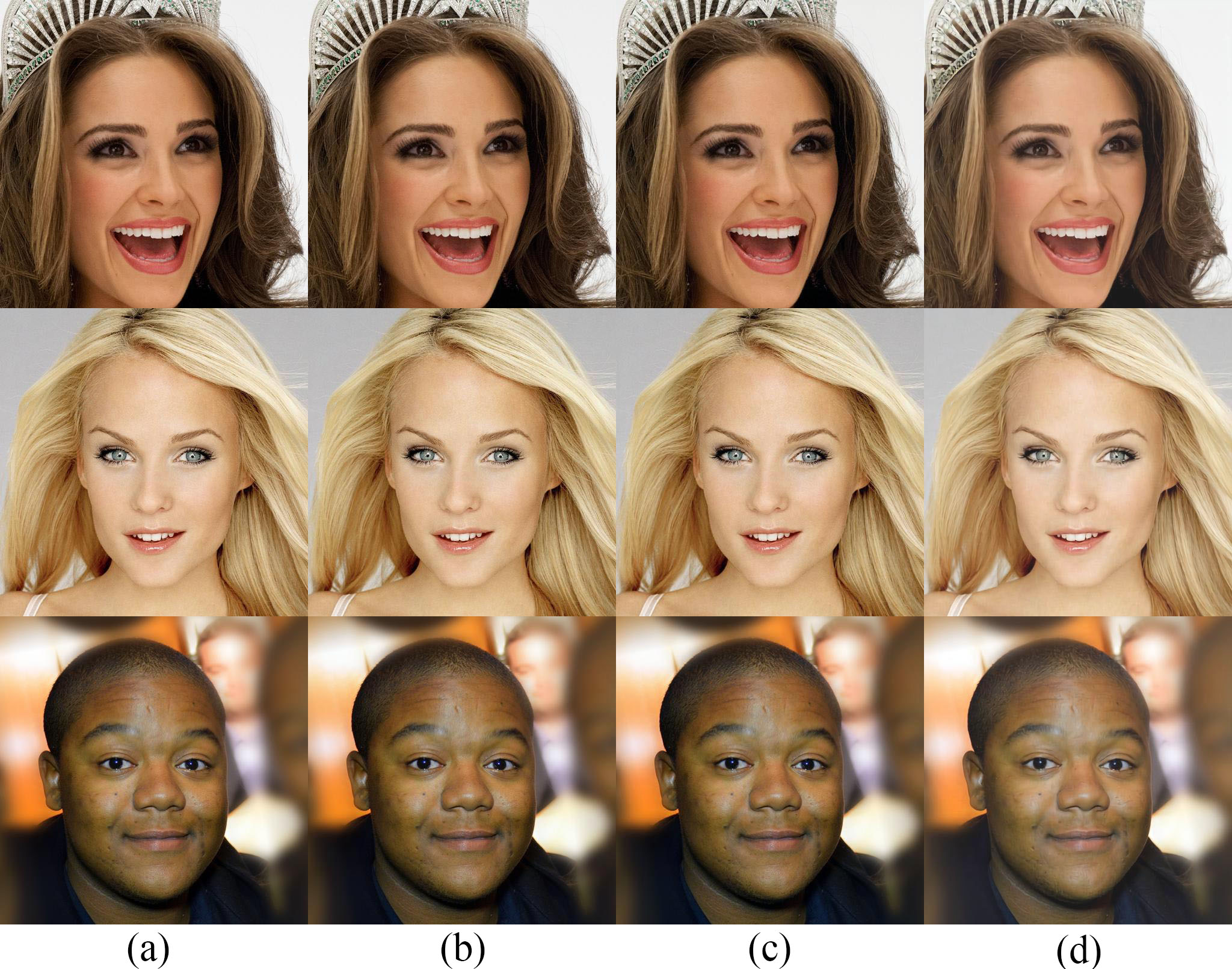}
\end{center}
\caption{Reconstructed images using different code sizes. (a) Input $512\times 512$ images. Reconstructed images using (b) $256\times 256\times 16$ (c) $128\times 128\times 16$ (d) $64\times 64\times 16$ code size. The reconstructed images become smooth when using $64\times 64\times 16$ code size.}
\label{fig:reconstruct}
\end{figure}

\paragraph{Code size.}
We propose to transform large images to small latent codes by training encoder and decoder networks. It is inevitable that part of the information in the original images will be lost after the encoder and decoder process. The size of the latent code determines the reconstruction errors. Smaller latent codes would make the code generation process faster, but decoding from the codes will be less accurate. For images with resolution $h\times w$, we try to encoder them to three different sizes of latent codes, $h/2\times w/2\times 16$, $h/4\times w/4\times 16$ and $h/8\times w/8\times 16$. The networks are all trained using $128\times 128$ images. The reconstruction mean square errors corresponding to different code sizes and resolutions are listed in Table~\ref{tab:reconstruction}. Smaller code size leads to larger reconstruction error. We also notice that the reconstruction error decreases when the input resolution increases. Even though the encoder and decoder networks are all trained using $128\times 128$ images, it works even better for higher resolutions. In Figure~\ref{fig:reconstruct}, we display the reconstruction $512\times 512$ images from three different code sizes. The reconstructed images from code size $256\times 256\times 16$ and $128\times 128\times 16$ look nearly the same with the input images. When using code size $64\times 64\times 16$, the reconstructed images miss some details and look smooth. In our experiments, we adopt $h/4\times w/4\times 16$ for the best balance of the reconstruction accuracy and the size of latent codes.

\begin{table*}
\begin{center}
\caption{Reconstruction mean square error (MSE) of the encoder and decoder networks with different code sizes and input resolutions.}
\label{tab:reconstruction}
\begin{tabular}{|c|c|c|c|c|}
\hline
\multirow{2}*{Code Size} & \multicolumn{4}{|c|}{Input Resolution} \\
\cline{2-5}
 & $128\times 128$ & $256\times 256$ & $512\times 512$ & $1024\times 1024$\\
\hline\hline
$h/2\times w/2\times 16$ & 2.7e-4 & 2.0e-4 & 1.5e-4 & 1.1e-4\\
$h/4\times w/4\times 16$ & 1.3e-3 & 1.0e-3 & 7.5e-4 & 3.5e-4\\
$h/8\times w/8\times 16$ & 3.4e-3 & 2.8e-3 & 2.0e-3 & 8.8e-4\\
\hline
\end{tabular}
\end{center}
\end{table*}

\paragraph{Training encoder and decoder networks.}
Though the proposed framework is for accelerating high resolution image synthesis, the encoder and decoder networks can be trained using low resolution images. In our experiments, the encoder and decoder networks are trained using $128\times 128$ images for efficiency consideration. Note that the encoder and decoder networks can be applied for any other resolution as both of them are fully convolutional neural networks. As shown in Table~\ref{tab:reconstruction}, the reconstruction error is even lower for high resolution input than low resolution input, though the same encoder and decoder networks are used. Therefore, the encoder and decoder networks only need to be trained once using low resolution images, no matter images at what resolutions we want to generate.

\paragraph{Normalization techniques.}
Normalization techniques are not used in our encoder and decoder networks as we find that they can be trained easily and stably without normalization techniques. In the generator network, we use pixel normalization \cite{karras2017progressive} after convolutional operations. As pointed out by \cite{karras2017progressive}, we also observe in our experiments that pixel normalization does not seem to change the results much. We add pixel normalization simply to prevent possible escalation of signal magnitudes during training.

\section{Experiments}
To measure whether the proposed framework is able to accelerate the training of high resolution image generation, we build a traditional image generation network and accelerate this network using the proposed framework.  We mainly compare the results of the networks before and after using the proposed acceleration framework, in terms of image quality, quantitative measurement and training speed.
\subsection{Experimental Settings}
The image generation network is nearly the same with Figure~\ref{fig:network}(b), except that the dimension of the output from the generator and the input to the discriminator is different, which becomes the size of images. For example, to generate $512\times 512$ images, the output of the image generator is $512\times 512\times 3$. Applying the proposed acceleration framework to this image generation network, we train a code generation network using exactly the same structure but with output size $128\times 128\times 16$, which is the size of latent codes. By comparison, the image generator and discriminator networks contain more layers than the code generator and discriminator. Whereas, the general structures of both methods are identical. The layer details of the network corresponding to resolution $256\times 256$ before and after acceleration is in Table~\ref{tab:layers}.

\begin{table}
\begin{center}
\caption{Feature map sizes corresponding to resolution $256\times 256$ before and after acceleration.}
\label{tab:layers}
\begin{tabular}{|c|c|c|}
\hline
Layers & Before acceleration  & After acceleration\\
\hline\hline
\multicolumn{3}{|c|}{Generator} \\
\hline
dense layer & $4096$ & $4096$ \\
conv & \tabincell{c}{$4\times4\times256$\\...\\$256\times256\times64$}   & \tabincell{c}{$4\times4\times256$\\...\\$64\times64\times128$}\\
output & $256\times256\times3$ & $64\times64\times16$ \\
\hline
\multicolumn{3}{|c|}{Discriminator} \\
\hline
conv & \tabincell{c}{$256\times256\times64$\\...\\$4\times4\times256$}   & \tabincell{c}{$64\times64\times128$\\...\\$4\times4\times256$}\\
dense layer & $256$ & $256$ \\
output & $1$ & $1$ \\
\hline
\end{tabular}
\end{center}
\end{table}

The dataset we use is celeba-hq dataset \cite{karras2017progressive}, which contains 30000 faces at resolution $1024\times 1024$. We resize images to $256\times 256$ and $512\times 512$ and learn to generate images at three resolutions respectively, i.e. $256\times 256$, $512\times 512$ and $1024\times 1024$. We display the generated images before and after acceleration, and evaluate the generated samples quantitatively by calculating the Fr\'echet Inception distance (FID) \cite{heusel2017gans}. We do not calculate the Inception score \cite{salimans2016improved} as we only generate the faces using celeba-hq dataset, not images in multiple categories. In addition, FID is considered to be consistent with human evaluation in terms of measuring the realism and variation of the generated images \cite{zhang2018self}.

All experiments are run on one NVIDIA P100 graphic card with 16GB memory, and our CPU is Intel Xeon E5-2682 V4 @ 2.5GHz.

\subsection{Training Speed}
We analyze that after acceleration, the generative network only needs to generate small latent codes instead of large images. The width of the codes in our experiments is one fourth of the width of the original images. As a result, there are less layers and smaller feature maps in the code generator and discriminator networks after acceleration than before. Notice that the proposed framework has to train additional encoder and decoder networks. As described in previous section, we can train these two networks using $128\times 128$ images, which converges relatively fast. It only take about 4 hours and 30 minutes to train the encoder and decoder networks, which is much less than training high resolution image generator and discriminator networks. In addition, for any other resolutions or datasets, we only have to train the encoder and decoder once. Therefore, we just compare the speed of training the generative networks.

The training speed of running one epoch (feeding in 30000 images) before and after acceleration is listed in Table~\ref{tab:speed}. We can see that, for all three resolutions, the training speed is highly accelerated by more than two times. For resolution $1024\times 1024$, the training speed after acceleration is 5 times faster than before, which makes it possible to train $1024\times 1024$ image generator in 3 days using only one P100 graphic card.

\begin{table}
\begin{center}
\caption{The running time of training one epoch (feeding in 30000 images).}
\label{tab:speed}
\begin{tabular}{|l|c|c|}
\hline
Resolution & Before acceleration & After acceleration\\
\hline\hline
$256\times 256$ & 30 minutes & \textbf{8 minutes} \\
$512\times 512$ &  56 minutes & \textbf{22 minutes} \\
$1024\times 1024$ & 225 minutes & \textbf{45 minutes}\\
\hline
\end{tabular}
\end{center}
\end{table}

\begin{figure}[h]
\centering
   \begin{subfigure}[b]{225pt}
   \includegraphics[width=225pt]{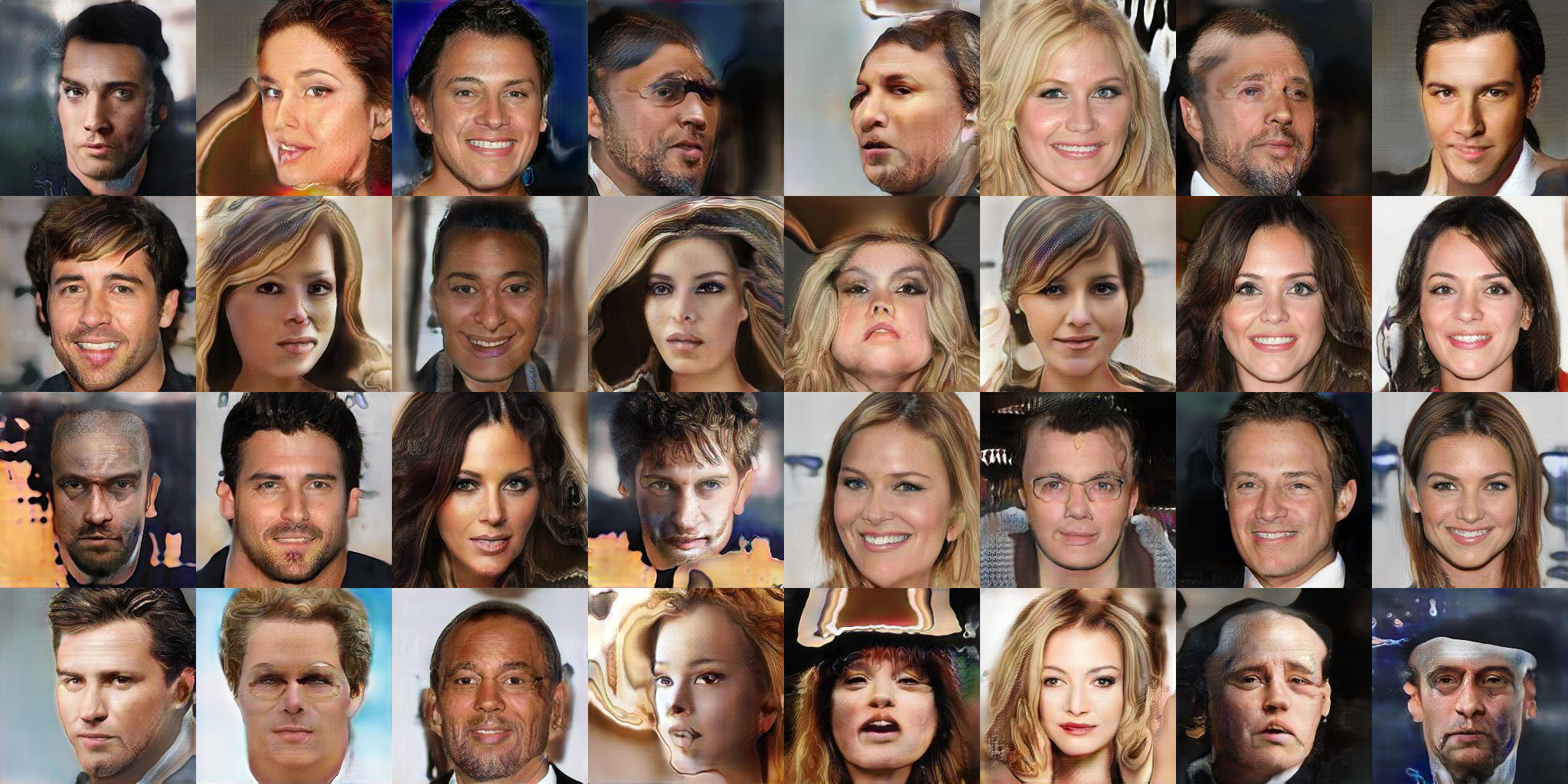}
   \caption{Before acceleration (FID=23.62)}
   \end{subfigure}
   \begin{subfigure}[b]{225pt}
   \includegraphics[width=225pt]{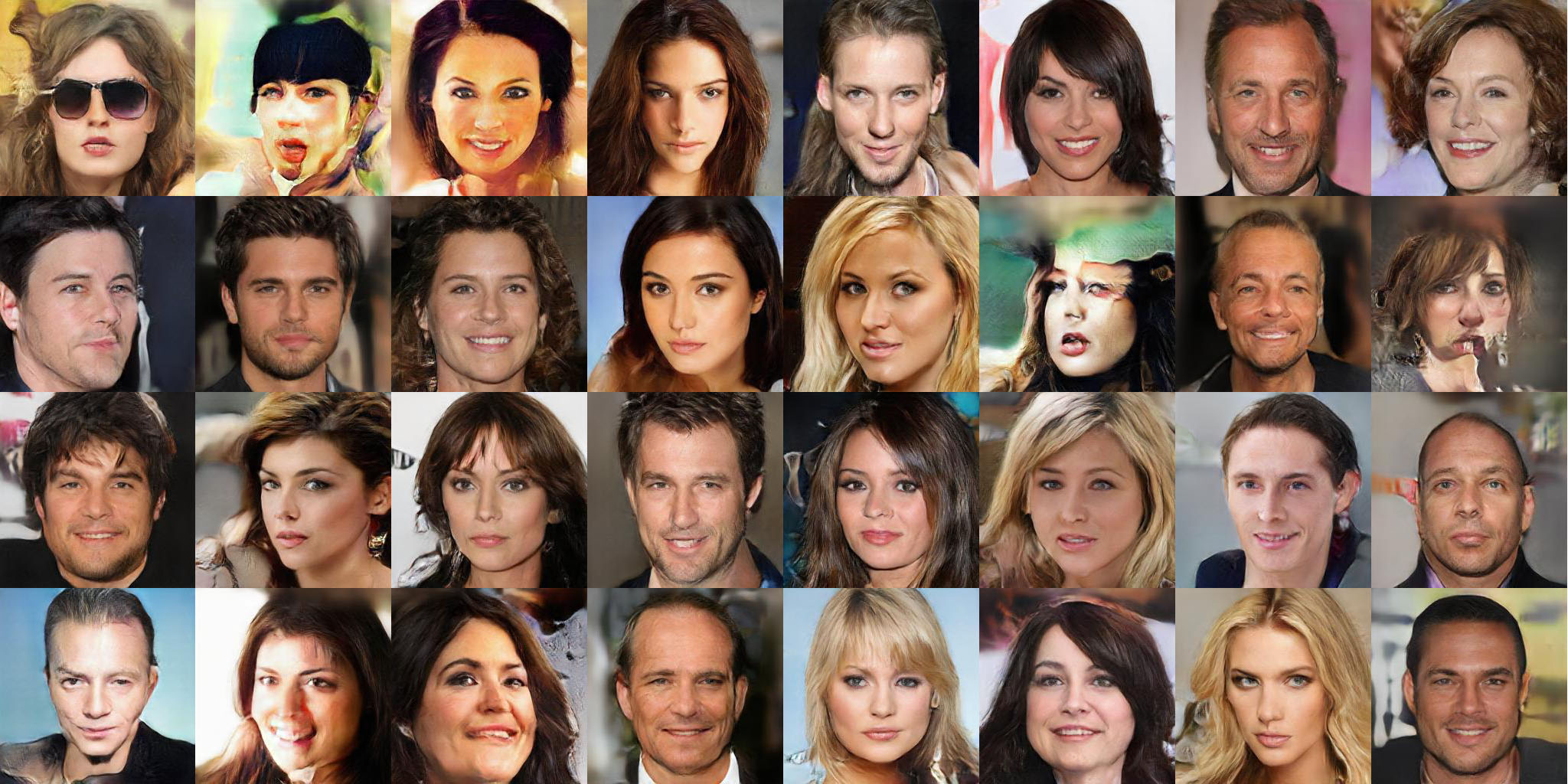}
   \caption{After acceleration (FID=20.78)}
   \end{subfigure}
\caption{Randomly generated $256\times 256$ images without manual selection (a) before and (b) after acceleration. The quality of the generated samples after acceleration are comparable with those before acceleration.}
\label{fig:256}
\end{figure}

\begin{figure}[h]
\centering
   \begin{subfigure}[b]{225pt}
   \includegraphics[width=225pt]{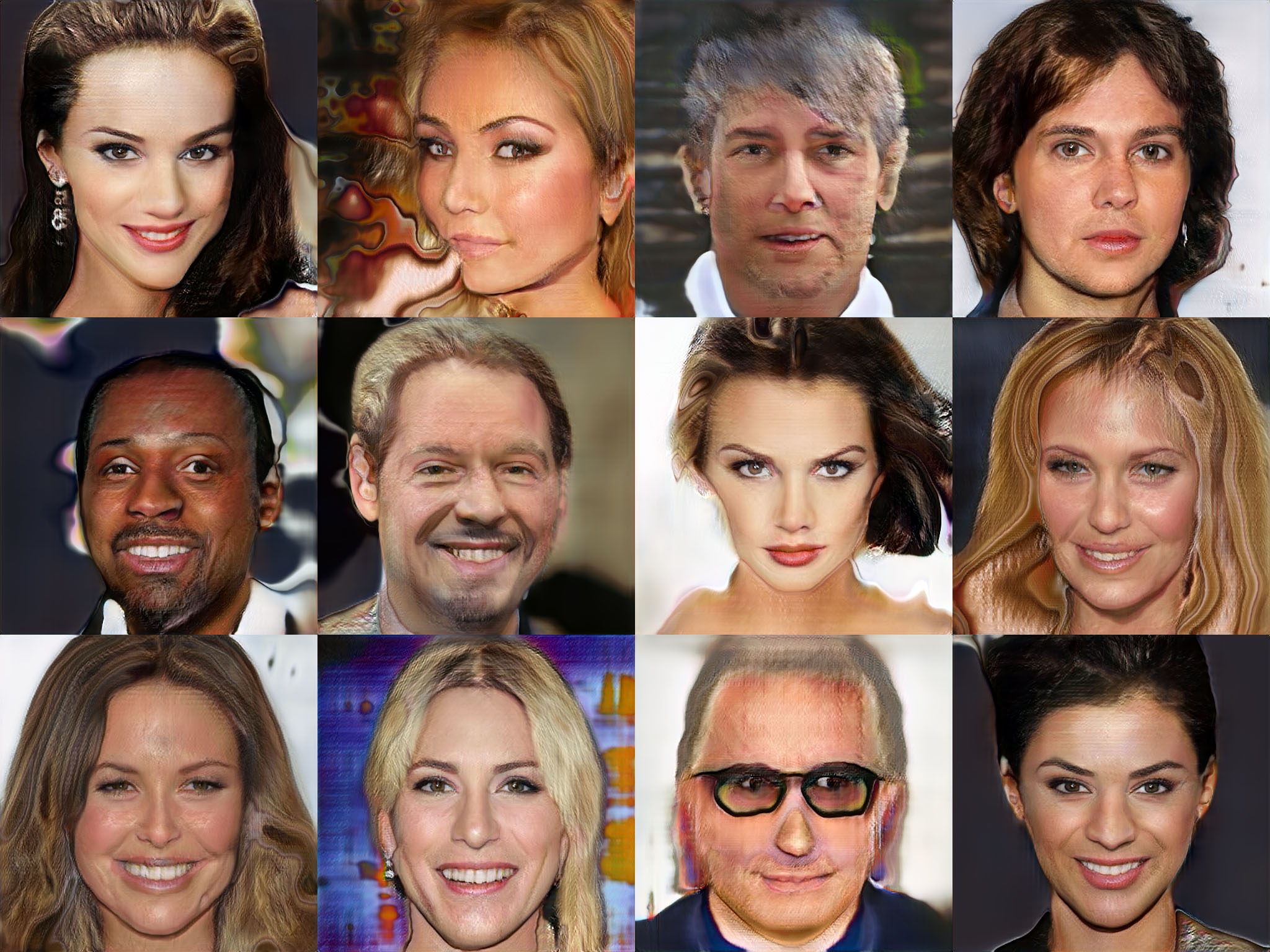}
   \caption{Before acceleration (FID=30.43)}
   \end{subfigure}
   \begin{subfigure}[b]{225pt}
   \includegraphics[width=225pt]{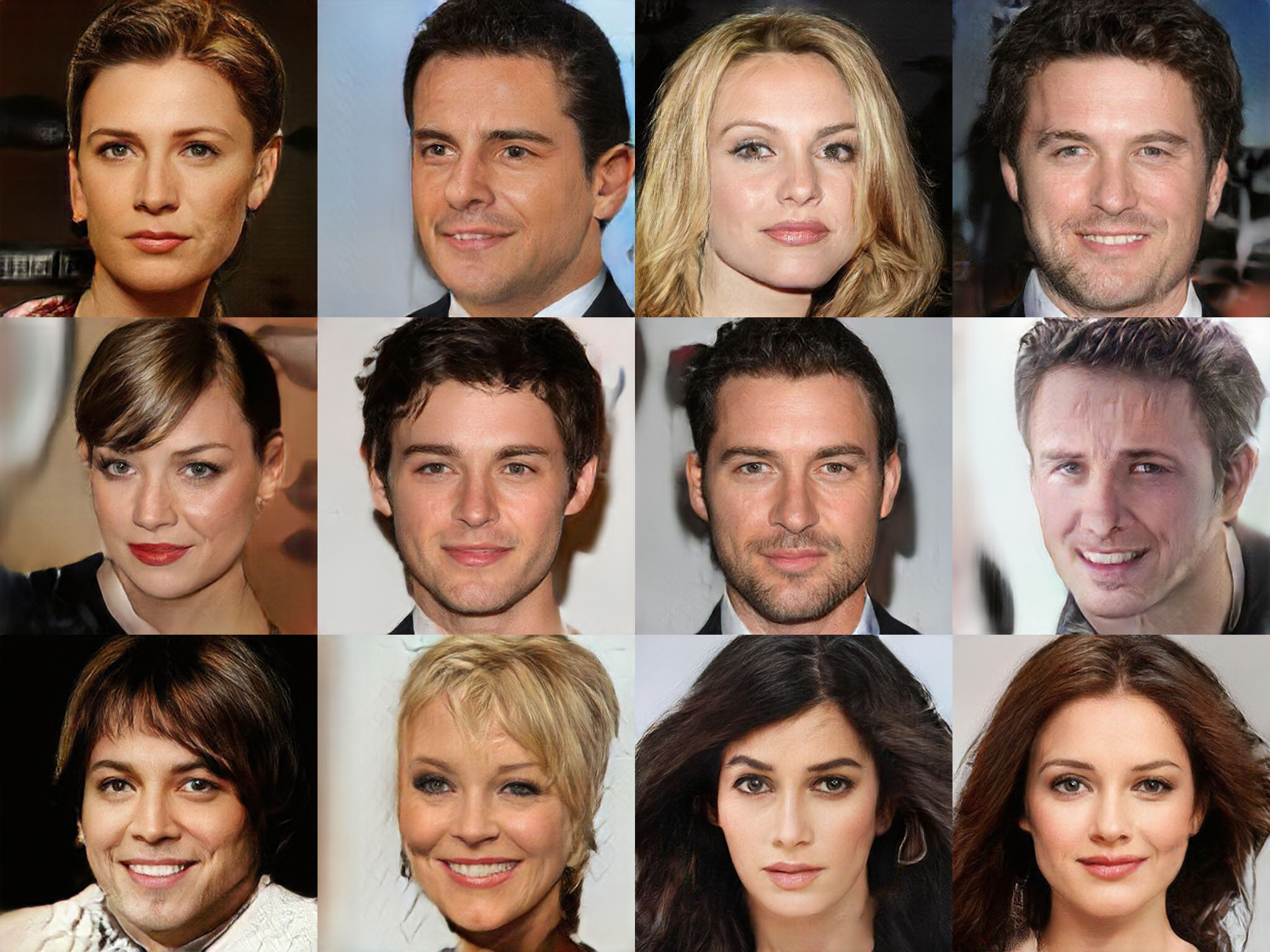}
   \caption{After acceleration (FID=14.72)}
   \end{subfigure}
\caption{Randomly generated $512\times 512$ images without manual selection (a) before and (b) after acceleration. The quality of the generated samples after acceleration looks better and with less artifacts those before acceleration.}
\label{fig:512}
\end{figure}

\subsection{Qualitative Evaluation}
\begin{figure*}
\centering
   \begin{subfigure}[b]{450pt}
   \includegraphics[width=450pt]{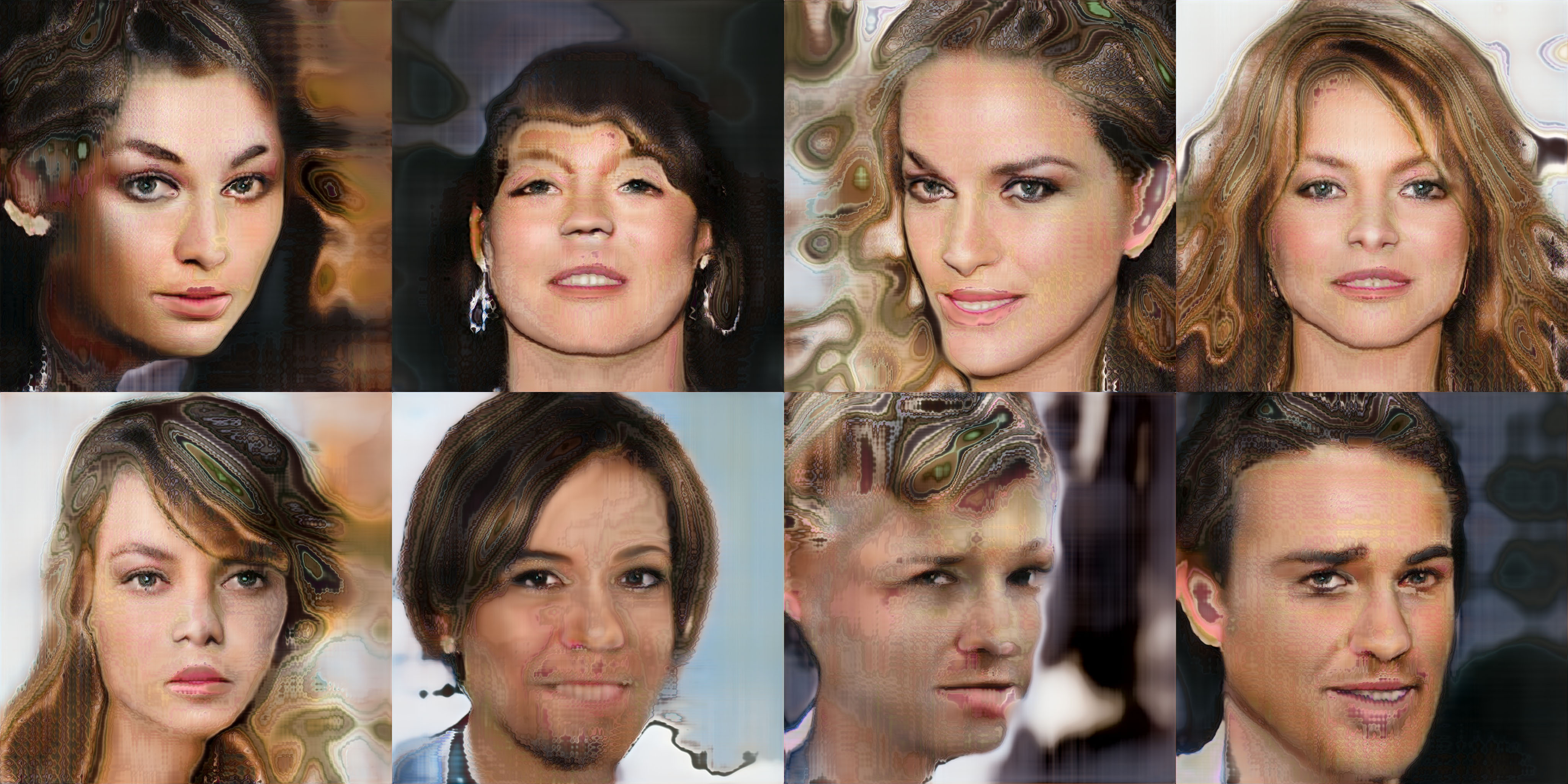}
   \caption{Before acceleration (FID=54.83)}
   \end{subfigure}
   \begin{subfigure}[b]{450pt}
   \includegraphics[width=450pt]{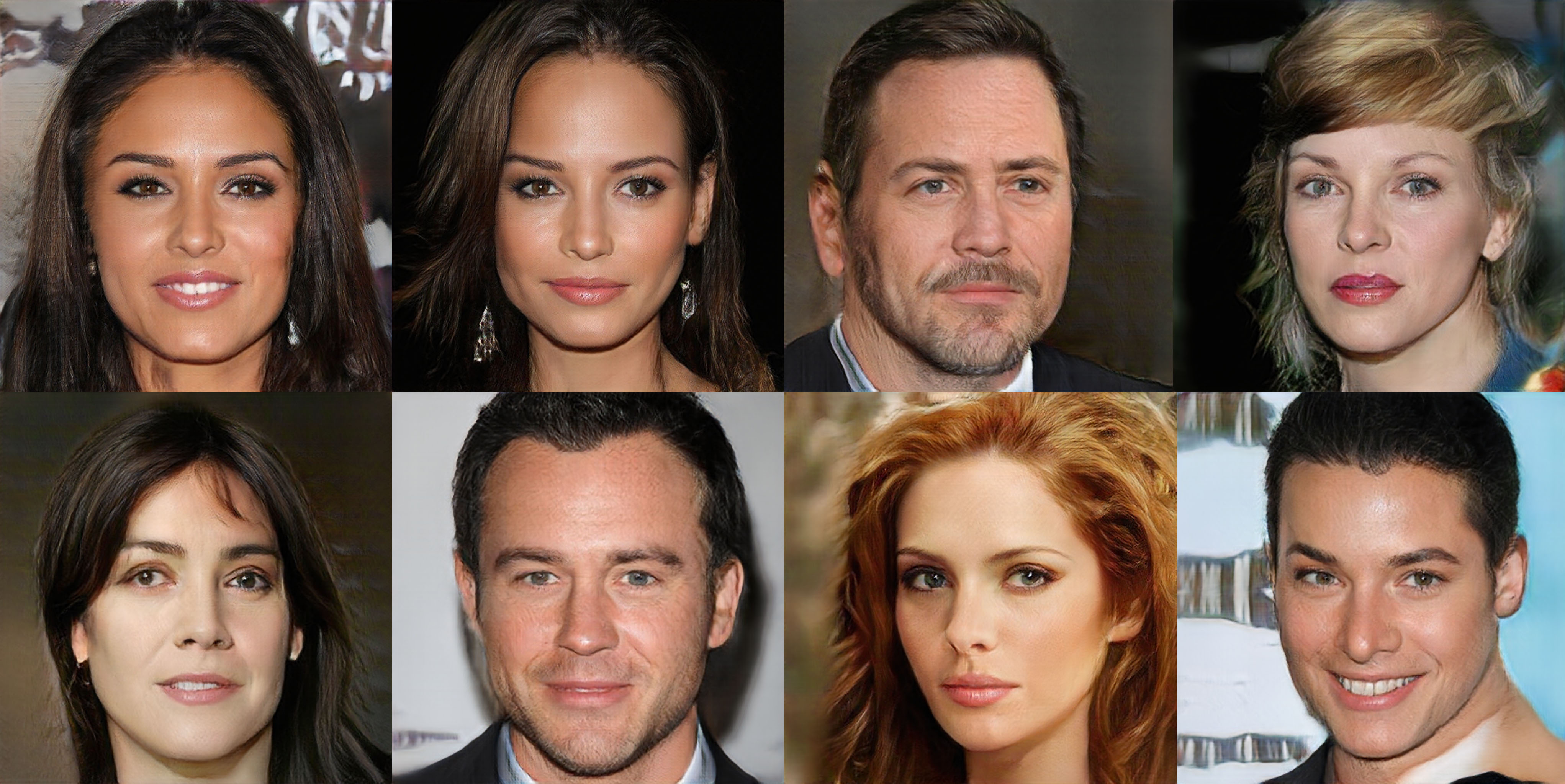}
   \caption{After acceleration (FID=14.80)}
   \end{subfigure}
\caption{Randomly generated $1024\times 1024$ images without manual selection (a) before and (b) after acceleration. The quality of the generated samples after acceleration is significantly better than before acceleration.}
\label{fig:1024}
\end{figure*}

A good acceleration method should increase the speed and keep the quality. We first qualitatively analyze of the generated samples before and after acceleration. High resolution images at $256\times 256$, $512\times 512$ and $1024\times 1024$ are generated respectively, which are displayed in Figure~\ref{fig:256}, \ref{fig:512} and \ref{fig:1024}. For resolution $256\times 256$ and $512\times 512$, the traditional network is able to generate reasonable and good results. Whereas, when it comes to $1024\times 1024$ resolution, the generated samples contain lots of artifacts. We consider that the large number of parameters and size of feature maps increase the difficulty of network training significantly at $1024\times 1024$ resolution. In contrast, after being accelerated by the proposed framework, the network is able to generate good samples at all resolutions. In addition, the general appearance of the samples after acceleration look more natural with less artifacts than before.

The results show that the proposed acceleration framework does not lower the quality of the original network. On the contrary, the quality of generated samples increases after acceleration. Our framework converts large image generation to small code generation, which enables the networks to learn less parameters and easier to converge. Therefore, after acceleration, the network shows better stability to generate high resolution images with satisfying image quality.

%

\begin{table}
\begin{center}
\caption{FID before and after acceleration.}
\label{tab:quantitative}
\begin{tabular}{|l|c|c|}
\hline
Resolution & Before acceleration & After acceleration\\
\hline\hline
$256\times 256$ & 23.62 & \textbf{20.78} \\
$512\times 512$ &  30.43 & \textbf{14.72} \\
$1024\times 1024$ & 54.83 & \textbf{14.80} \\
\hline
\end{tabular}
\end{center}
\end{table}

\subsection{Quantitative Evaluation}
We further evaluate the generated samples quantitatively. We randomly generate 50k samples after training and FIDs corresponding to different resolutions before and after acceleration are calculated. From Table~\ref{tab:quantitative}, we can see that after using the proposed acceleration method, the FID decreases. Before acceleration, the network is able to generate good samples with relatively low FIDs at resolution $256\times 256$ and $512\times 512$, but fails at higher resolutions. FIDs are very large when generating $1024\times 1024$ samples. The proposed method decreases the FIDs at all three resolutions. For resolution $512\times 512$ and $1024\times 1024$, the improvements are significant. In conclusion, the quality of generated samples is well improved after using the proposed acceleration framework, which is in accordance with the qualitative measurement.

\subsection{Other Datasets}
\begin{figure}
\centering
   \begin{subfigure}[b]{225pt}
   \includegraphics[width=225pt]{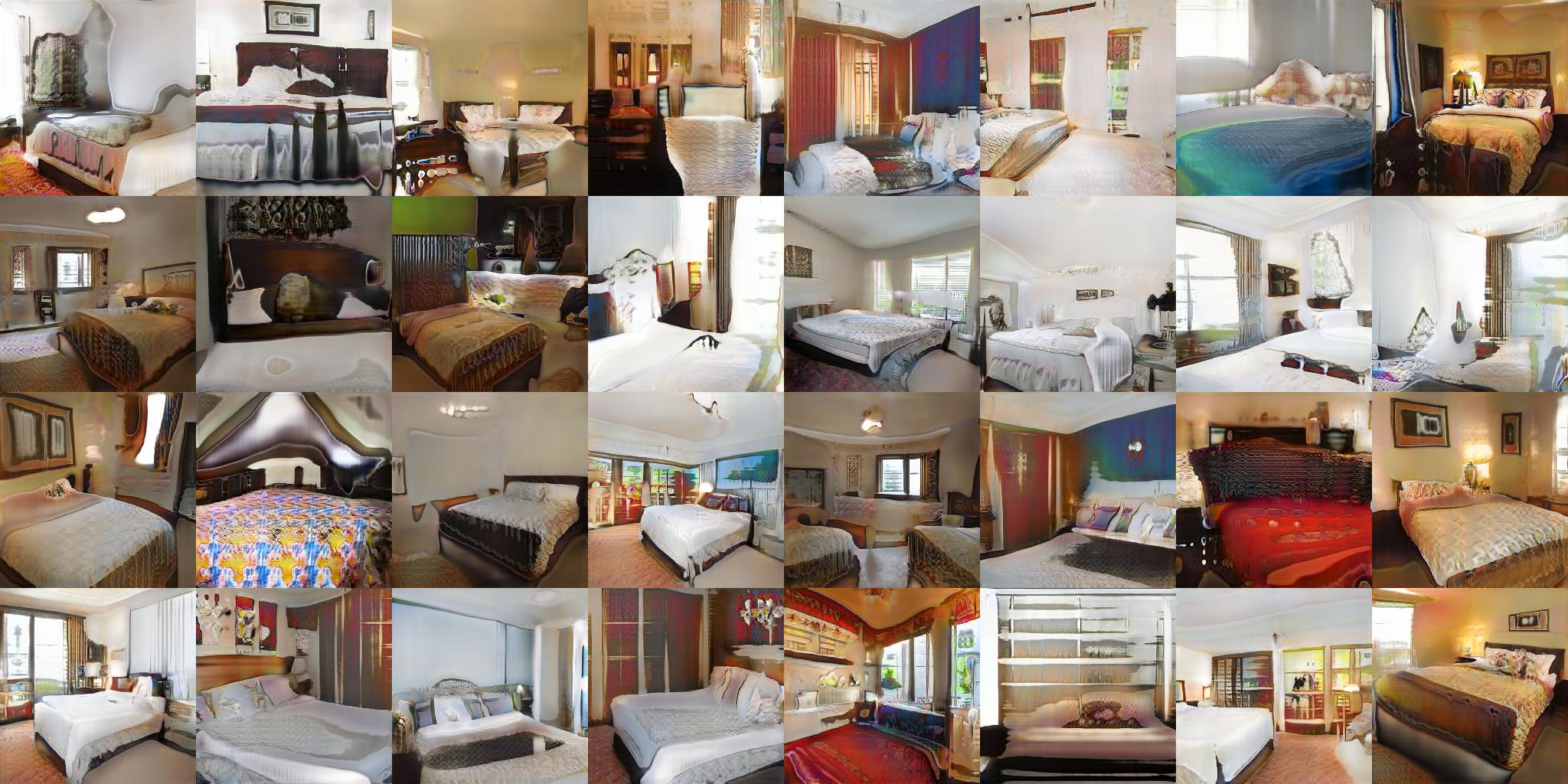}
   \caption{Before acceleration (FID=38.32)}
   \end{subfigure}
   \begin{subfigure}[b]{225pt}
   \includegraphics[width=225pt]{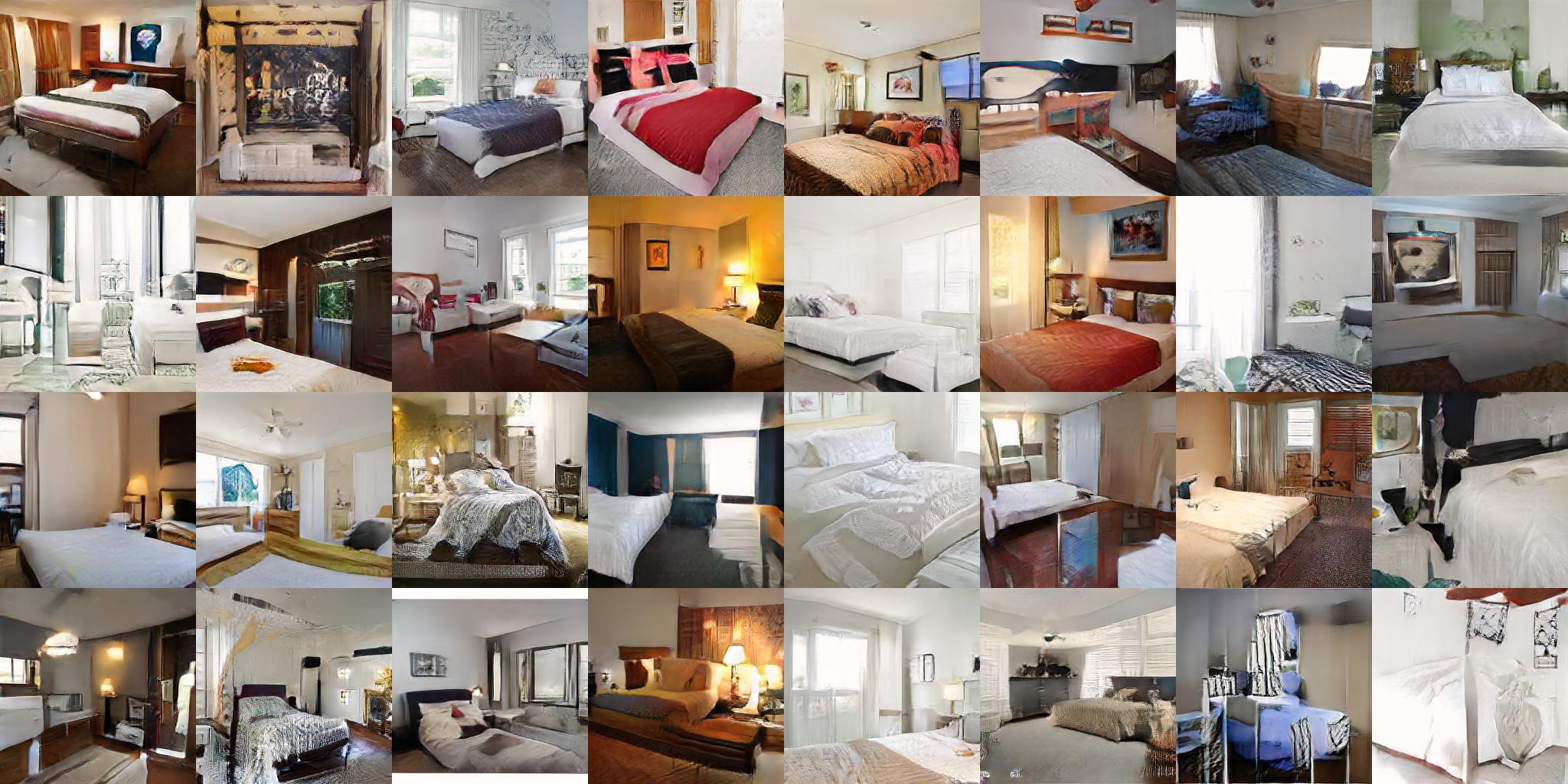}
   \caption{After acceleration (FID=17.90)}
   \end{subfigure}
\caption{Randomly generated $256\times 256$ samples using lsun bedroom dataset without manual selection (a) before and (b) after acceleration. The quality of the generated samples is improved after acceleration.}
\label{fig:bedroom}
\end{figure}

\begin{figure}
\centering
   \begin{subfigure}[b]{225pt}
   \includegraphics[width=225pt]{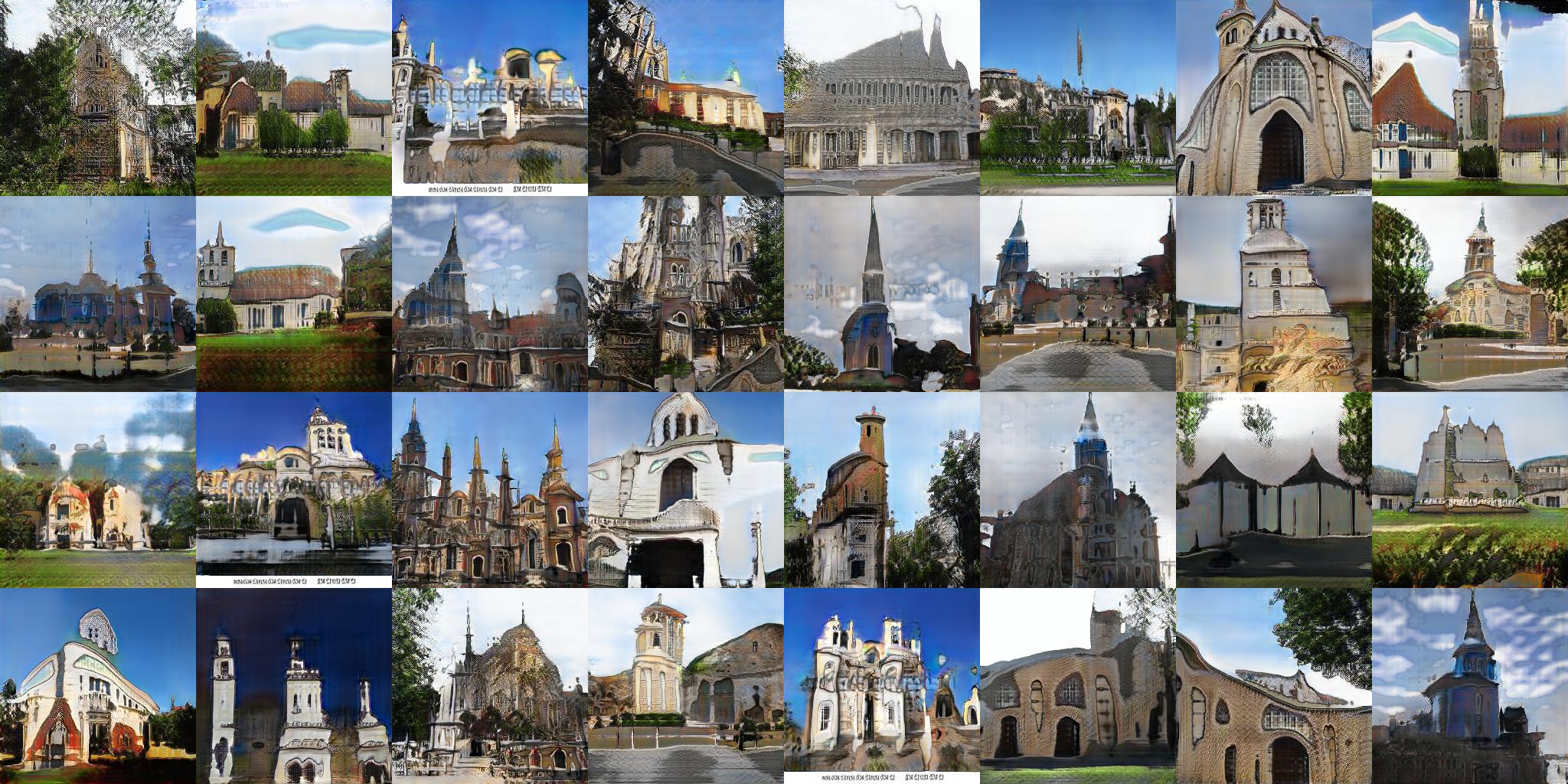}
   \caption{Before acceleration (FID=30.99)}
   \end{subfigure}
   \begin{subfigure}[b]{225pt}
   \includegraphics[width=225pt]{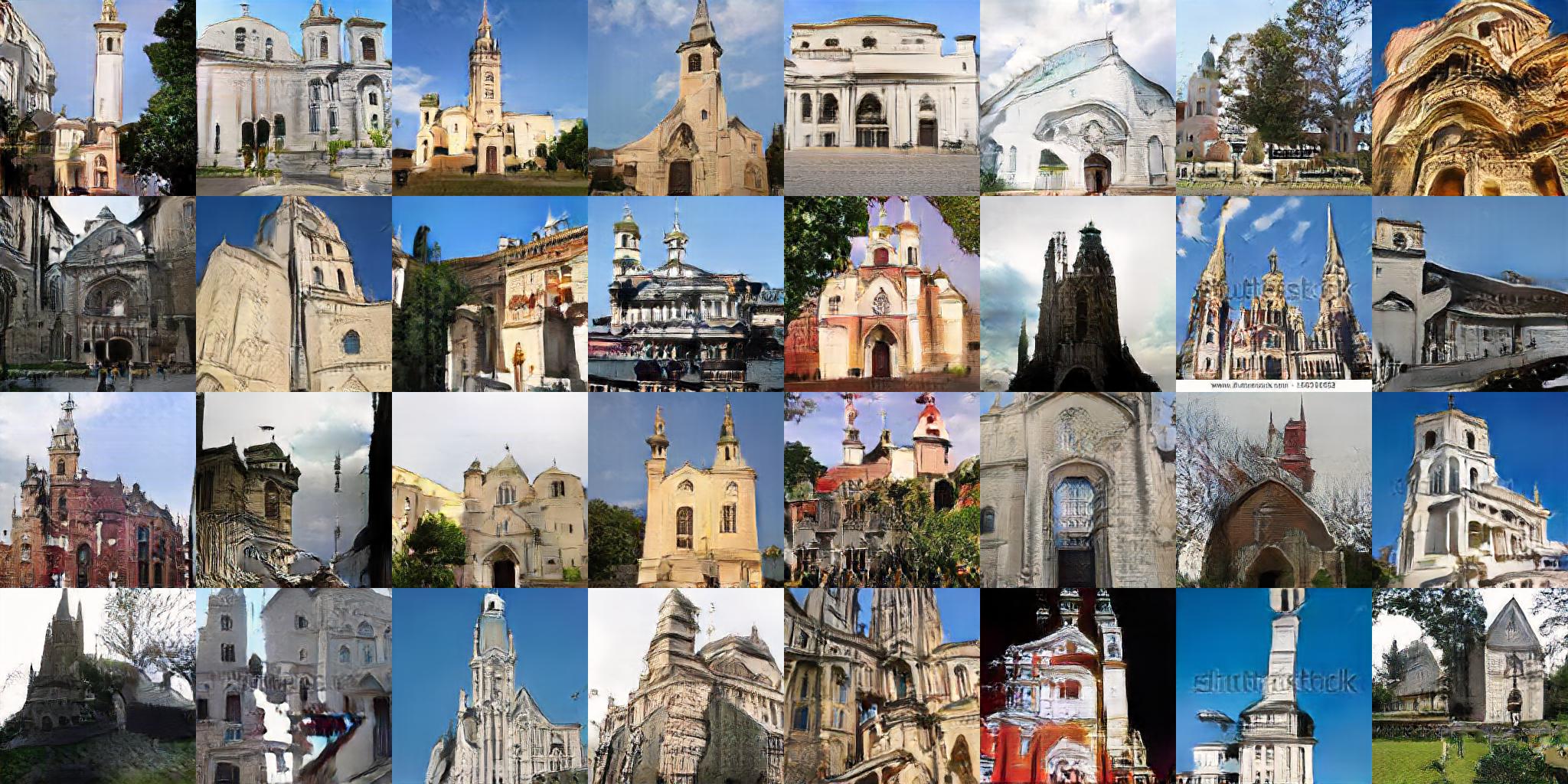}
   \caption{After acceleration (FID=21.91)}
   \end{subfigure}
\caption{Randomly generated $256\times 256$ samples using lsun church dataset without manual selection (a) before and (b) after acceleration. The quality of the generated samples is improved after acceleration.}
\label{fig:church}
\end{figure}

We further test the proposed method on other datasets. Lsun datasets \cite{yu15lsun} at resolution $256\times 256$ are used to train the networks. As mentioned in the theory part, the encoder and decoder networks only need to be trained once. Thus, we do not retrain the encoder and decoder networks using these specific datasets. Instead, we use the encoder and decoder networks that are trained on the celeba-hq dataset directly. Even though, the proposed framework is able to generate reasonable samples as displayed in Figure~\ref{fig:bedroom} and \ref{fig:church}. In addition, we calculates the FIDs in Table~\ref{tab:category}. The proposed acceleration method improves the quality on all datasets as well.

\begin{table*}
\begin{center}
\caption{FID of LSUN datasets before and after acceleration.}
\label{tab:category}
\begin{tabular}{|c|c|c|c|c|c|c|}
\hline
LSUN Dataset  & cat & airplane & bus & bird & bedroom & church\\
\hline\hline
before acceleration & 66.90  & 72.53 & 25.75 & 62.44 & 38.32 & 30.99 \\
after acceleration &  \textbf{34.25}  & \textbf{25.78} & \textbf{15.29} & \textbf{28.42} & \textbf{17.90} & \textbf{21.91}\\
\hline
\end{tabular}
\end{center}
\end{table*}

\begin{table}[t]
\begin{center}
\caption{FID and the running time of training one epoch corresponding to different settings.}
\label{tab:code}
\begin{tabular}{|c|c|c|c|c|}
\hline
Settings  & FID & Running time\\
\hline\hline
without acceleration & 54.83 & 225 minutes \\
$h/4 \times w/4 \times 16$ &  \textbf{14.80} & 45 minutes \\
$h/8 \times w/8 \times 16$ & 17.71 & 19 minutes \\
$h/16 \times w/16 \times 16$ & 22.12 & \textbf{9 minutes} \\
\hline
\end{tabular}
\end{center}
\end{table}

\section{Limitations and Discussions}
The target of the proposed method is to accelerate the training process of high resolution image generation without lowering the quality. Therefore, we did not spend much time building complex networks. We build a relatively simple generative network and test the results before and after using the proposed acceleration framework. The proposed method enables generating promising samples at high resolutions within short training time. Whereas, the generated images are not as good as recent advanced image generation structures such as \cite{karras2018style}. More complicated generative network should be used to reach the quality of the state-of-art methods in our following work.

In experiments, we adopt $h/4\times w/4 \times 16$ code size. In fact, the code can be much smaller. We adopt this size for the best quality of generated images. For resolution $1024\times 1024$, we test smaller code sizes in Table~\ref{tab:code}. Even for $h/16 \times w/16 \times 16$ code size, the FID is much less after acceleration and the speed is the fastest.

\begin{figure}[t]
\begin{center}
  \includegraphics[width=230pt]{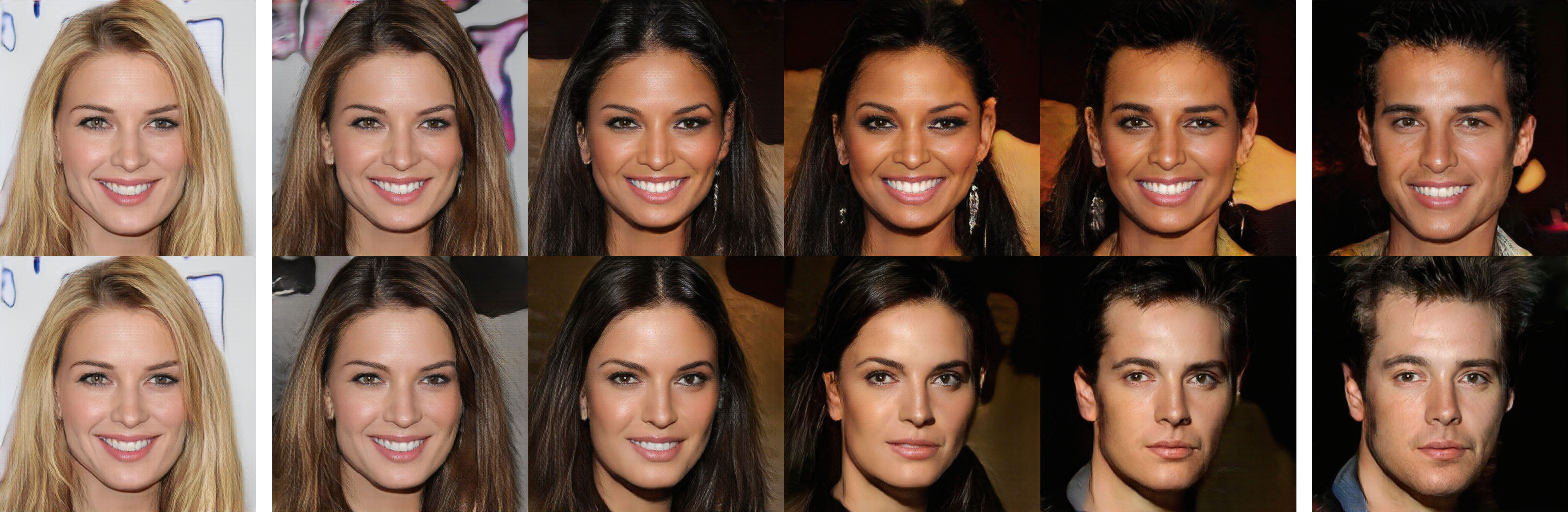}
\end{center}
\caption{Input noise latent space interpolation results.}
\label{fig:interpolation}
\end{figure}

The input noise latent space interpolation results are displayed in Figure~\ref{fig:interpolation}. The generated images change smoothly, which shows that the proposed frame-work does not just memorize training samples.

\section{Conclusion}
We propose an acceleration framework for high resolution images generation in this paper. Encoder and decoder networks are first trained to transform large images to small latent codes. Then the code generation networks learn to generate small codes in latent space.The training process is highly accelerated and the network stability is well improved. Experimental results show that the proposed framework makes the training speed times faster, and improves the generated image quality as well. The proposed acceleration framework makes it possible to generate satisfying high resolution images using less training time with limited hardware resource.

{\small
\bibliographystyle{ieee}
\bibliography{fastGAN_arxiv}

\begin{thebibliography}{10}\itemsep=-1pt

\bibitem{arjovsky2017wasserstein}
M.~Arjovsky, S.~Chintala, and L.~Bottou.
\newblock Wasserstein generative adversarial networks.
\newblock In {\em International Conference on Machine Learning}, pages
  214--223, 2017.

\bibitem{brock2018large}
A.~Brock, J.~Donahue, and K.~Simonyan.
\newblock Large scale gan training for high fidelity natural image synthesis.
\newblock {\em arXiv preprint arXiv:1809.11096}, 2018.

\bibitem{goodfellow2014generative}
I.~Goodfellow, J.~Pouget-Abadie, M.~Mirza, B.~Xu, D.~Warde-Farley, S.~Ozair,
  A.~Courville, and Y.~Bengio.
\newblock Generative adversarial nets.
\newblock In {\em Advances in neural information processing systems}, pages
  2672--2680, 2014.

\bibitem{gulrajani2017improved}
I.~Gulrajani, F.~Ahmed, M.~Arjovsky, V.~Dumoulin, and A.~C. Courville.
\newblock Improved training of wasserstein gans.
\newblock In {\em Advances in Neural Information Processing Systems}, pages
  5767--5777, 2017.

\bibitem{heusel2017gans}
M.~Heusel, H.~Ramsauer, T.~Unterthiner, B.~Nessler, and S.~Hochreiter.
\newblock Gans trained by a two time-scale update rule converge to a local nash
  equilibrium.
\newblock In {\em Advances in Neural Information Processing Systems}, pages
  6626--6637, 2017.

\bibitem{karras2017progressive}
T.~Karras, T.~Aila, S.~Laine, and J.~Lehtinen.
\newblock Progressive growing of gans for improved quality, stability, and
  variation.
\newblock {\em arXiv preprint arXiv:1710.10196}, 2017.

\bibitem{karras2018style}
T.~Karras, S.~Laine, and T.~Aila.
\newblock A style-based generator architecture for generative adversarial
  networks.
\newblock {\em arXiv preprint arXiv:1812.04948}, 2018.

\bibitem{kingma2013auto}
D.~P. Kingma and M.~Welling.
\newblock Auto-encoding variational bayes.
\newblock {\em arXiv preprint arXiv:1312.6114}, 2013.

\bibitem{mescheder2018training}
L.~Mescheder, A.~Geiger, and S.~Nowozin.
\newblock Which training methods for gans do actually converge?
\newblock {\em arXiv preprint arXiv:1801.04406}, 2018.

\bibitem{miyato2018spectral}
T.~Miyato, T.~Kataoka, M.~Koyama, and Y.~Yoshida.
\newblock Spectral normalization for generative adversarial networks.
\newblock {\em arXiv preprint arXiv:1802.05957}, 2018.

\bibitem{odena2017conditional}
A.~Odena, C.~Olah, and J.~Shlens.
\newblock Conditional image synthesis with auxiliary classifier gans.
\newblock In {\em Proceedings of the 34th International Conference on Machine
  Learning-Volume 70}, pages 2642--2651. JMLR. org, 2017.

\bibitem{salimans2016improved}
T.~Salimans, I.~Goodfellow, W.~Zaremba, V.~Cheung, A.~Radford, and X.~Chen.
\newblock Improved techniques for training gans.
\newblock In {\em Advances in neural information processing systems}, pages
  2234--2242, 2016.

\bibitem{smith2017improved}
E.~Smith and D.~Meger.
\newblock Improved adversarial systems for 3d object generation and
  reconstruction.
\newblock {\em arXiv preprint arXiv:1707.09557}, 2017.

\bibitem{wang2018high}
T.-C. Wang, M.-Y. Liu, J.-Y. Zhu, A.~Tao, J.~Kautz, and B.~Catanzaro.
\newblock High-resolution image synthesis and semantic manipulation with
  conditional gans.
\newblock In {\em Proceedings of the IEEE Conference on Computer Vision and
  Pattern Recognition}, pages 8798--8807, 2018.

\bibitem{wu2016learning}
J.~Wu, C.~Zhang, T.~Xue, B.~Freeman, and J.~Tenenbaum.
\newblock Learning a probabilistic latent space of object shapes via 3d
  generative-adversarial modeling.
\newblock In {\em Advances in neural information processing systems}, pages
  82--90, 2016.

\bibitem{yang20173d}
B.~Yang, H.~Wen, S.~Wang, R.~Clark, A.~Markham, and N.~Trigoni.
\newblock 3d object reconstruction from a single depth view with adversarial
  learning.
\newblock In {\em Proceedings of the IEEE International Conference on Computer
  Vision}, pages 679--688, 2017.

\bibitem{yu15lsun}
F.~Yu, Y.~Zhang, S.~Song, A.~Seff, and J.~Xiao.
\newblock Lsun: Construction of a large-scale image dataset using deep learning
  with humans in the loop.
\newblock {\em arXiv preprint arXiv:1506.03365}, 2015.

\bibitem{zhang2018self}
H.~Zhang, I.~Goodfellow, D.~Metaxas, and A.~Odena.
\newblock Self-attention generative adversarial networks.
\newblock {\em arXiv preprint arXiv:1805.08318}, 2018.

\bibitem{zhang2017stackgan++}
H.~Zhang, T.~Xu, H.~Li, S.~Zhang, X.~Wang, X.~Huang, and D.~Metaxas.
\newblock Stackgan++: Realistic image synthesis with stacked generative
  adversarial networks.
\newblock {\em arXiv preprint arXiv:1710.10916}, 2017.

\end{thebibliography}
}

\end{document}